\documentclass[aps,twocolumn,notitlepage]{revtex4-1}

\usepackage{algorithm,algpseudocode,amsfonts,amsmath,amssymb,amsthm,dsfont,graphicx,hyperref,hyphenat,lmodern,mathtools,microtype,multirow,tikz,xcolor}
\usepackage[UKenglish]{babel}
\usepackage[capitalize,nameinlink]{cleveref}
\usepackage[T1]{fontenc}
\usepackage[scaled=0.8]{FiraMono}
\usepackage[utf8]{inputenc}
\usepackage[binary-units]{siunitx}
\usepackage[caption=false]{subfig}

\definecolor{darkred}{rgb}{0.5,0,0}
\definecolor{darkgreen}{rgb}{0,0.5,0}
\definecolor{darkblue}{rgb}{0,0,0.5}
\hypersetup{colorlinks,breaklinks,linkcolor=darkred,citecolor=darkgreen,urlcolor=darkblue}

\usetikzlibrary{shapes.geometric,arrows,arrows.meta}
\tikzstyle{block} = [rectangle, draw, rounded corners, minimum width=2cm, minimum height=1cm, text centered, node distance=3cm, align=center, line width=0.5pt]
\tikzstyle{line} = [draw, -{Stealth[round, scale=1.4]}, line width=0.5pt]

\DeclareMathOperator*{\argmax}{arg\,max}
\DeclareMathOperator*{\argmin}{arg\,min}
\DeclareMathOperator{\sign}{sign}
\DeclareMathOperator{\Tr}{Tr}

\newtheorem{lemma}{Lemma}
\newtheorem{question}{Question}

\renewcommand{\leq}{\leqslant}
\renewcommand{\geq}{\geqslant}

\newcommand\hb[2]{\genfrac{}{}{0pt}{}{#1}{#2}}
\newcommand{\ba}{\mathbf{a}}
\newcommand{\bd}{\mathbf{d}}
\newcommand{\be}{\mathbf{e}}
\newcommand{\bp}{\mathbf{p}}
\newcommand{\br}{\mathbf{r}}
\newcommand{\bx}{\mathbf{x}}
\newcommand{\by}{\mathbf{y}}
\newcommand{\id}{\mathds{1}}
\newcommand{\innp}[1]{\langle #1 \rangle}
\newcommand{\ket}[1]{|#1\rangle}
\newcommand{\ketbra}[1]{|#1\rangle \langle#1|}
\newcommand{\mL}{\mathcal{L}}

\newcommand{\va}{\vec{a}}
\newcommand{\vb}{\vec{b}}
\newcommand{\vq}{\vec{q}}
\newcommand{\vu}{\vec{u}}
\newcommand{\vv}{\vec{v}}
\newcommand{\rvline}{\hspace*{-\arraycolsep}\vline\hspace*{-\arraycolsep}}

\graphicspath{{../arXiv_v3/}}

\begin{document}

\author{Sébastien Designolle}
\email{designolle@zib.de}
\author{Gabriele Iommazzo}
\author{Mathieu Besançon}
\author{Sebastian Knebel}
\author{Patrick Gelß}
\author{Sebastian Pokutta}
\affiliation{Zuse-Institut Berlin, Takustraße 7, 14195 Berlin, Germany}

\title{Improved local models and new Bell inequalities via Frank-Wolfe algorithms}
\date{5th October 2026}

\begin{abstract}
  In Bell scenarios with two outcomes per party, we algorithmically consider the two sides of the membership problem for the local polytope: constructing local models and deriving separating hyperplanes, that is, Bell inequalities.
  We take advantage of the recent developments in so-called Frank-Wolfe algorithms to significantly increase the convergence rate of existing methods.
  As an application, we study the threshold value for the nonlocality of two-qubit Werner states under projective measurements.
  Here, we improve on both the upper and lower bounds present in the literature.
  Importantly, our bounds are entirely analytical; moreover, they yield refined bounds on the value of the Grothendieck constant of order three: $1.4367\leq K_G(3)\leq1.4546$.
  We also demonstrate the efficiency of our approach in multipartite Bell scenarios, and present the first local models for all projective measurements with visibilities noticeably higher than the entanglement threshold.
  We make our entire code accessible as a Julia library called \texttt{BellPolytopes.jl}.
\end{abstract}

\maketitle

\textit{Introduction.---}
Long after the establishment of quantum mechanics, Bell uncovers in 1964 the concept of nonlocality~\cite{Bel64}.
Arguably one of the most striking features of the theory, this property makes it possible to distinguish correlations that can be obtained by classical or quantum means and since then has been extensively studied~\cite{BCP+14}.
Of particular interest is the question of the relation between this notion and the one of entanglement: although entanglement is clearly necessary to observe any bipartite nonlocality, asking whether it is sufficient is a delicate question.
For pure states, this is indeed the case~\cite{Gis91}, but this is not true for general states: in 1989, Werner exhibits mixed states that are entangled but nonetheless local~\cite{Wer89}.

More precisely, for a specific one-parameter family of states, by constructing an explicit local model recovering the correlations observed, he shows that the nonlocality threshold is different from the entanglement threshold.
However, although he computes the latter exactly, his proof only provides a bound on the former, sufficient to assess the above-mentioned phenomenon, but far from the actual value.
In the two-qubit case, the nonlocality witnessed by the Clauser--Horne--Shimony--Holt (CHSH) inequality for this family of states gives a bound in the opposite direction~\cite{CHSH69}.
The large interval between these two bounds remains untouched for almost two decades, until Ac\'in, Gisin, and Toner~\cite{AGT06} realise that, owing to a connection already seen by Tsirelson~\cite{Tsi87} to an equivalent mathematical problem (discussed below), an improved bound already existed~\cite{Kri79}, substantially reducing the gap.
Soon after, Ref.~\cite{Ver08} improves on the CHSH bound.

More recent works~\cite{BNV16,MW16} take numerical approaches, by relying on an optimisation algorithm by Gilbert~\cite{Gil66}, and on the known fact that the set of classical correlations is a polytope whose vertices are deterministic strategies (see, e.g.,~\cite{BCP+14}).
These works employ Gilbert's algorithm to approximate a quantum point, by minimising the distance to that point from this so-called local polytope.
This amounts to optimising linear approximations of a quadratic distance function, given by the local gradient, to iteratively move towards one of the polytope vertices.
The algorithm can converge to a facet without the need to compute the corresponding hyperplane.
New bounds have then been attained in~\cite{BNV16,DBV17,HQV+17} by combining this algorithm with other techniques.

\begin{table}[h]
  \centering
  \begin{tabular}{|c|c|rl|c|c|}
    \hline
                                                                              & $v_c^\mathrm{Wer}$       & \multicolumn{2}{c|}{Reference}                                  & \#Inputs                   & ~~Year~~               \\\hline
                                                                              & ~~0.7071~~               & \multicolumn{2}{c|}{CHSH~\cite{CHSH69}}                         & $2$                        & 1969                   \\\cline{2-6}
                                                                              & 0.7056                   & \multicolumn{2}{c|}{V\'ertesi~\cite{Ver08}}                     & $465$                      & 2008                   \\\cline{2-6}
                                                                              & 0.7054                   & \multicolumn{2}{c|}{Hua et al.~\cite{HLZ+15}}                   & $\infty$                   & 2015                   \\\cline{2-6}
                                                                              & 0.7012                   & \multicolumn{2}{c|}{Brierley et al.~\cite{BNV16}}               & $42$                       & 2016                   \\\cline{2-6}
                                                                              & 0.6964                   & \multicolumn{2}{c|}{~~Divi\'anszky et al.~\cite{DBV17}~~}       & $90$                       & 2017                   \\\cline{2-6}
    \parbox[t]{5mm}{\multirow{-6}{*}{\rotatebox[origin=c]{90}{Upper bounds}}} & 0.6961                   &                                         & $\!\!$\cref{eqn:vup}  & $97$                                 & \\\cline{1-2}\cline{5-5}
                                                                              & 0.6875                   & \multirow{-2}{*}{~~This work $\bigg\{$} & $\!\!$\cref{eqn:vlow} & $406\sim\infty$                      & \multirow{-2}{*}{2023} \\\cline{2-6}
                                                                              & 0.6829                   & \multicolumn{2}{c|}{Hirsch et al.~\cite{HQV+17}}                & ~$625\sim\infty$~          & 2017                   \\\cline{2-6}
                                                                              &                          & \multicolumn{2}{c|}{Ac\'in et al.~\cite{AGT06}}                 &                            & 2006                   \\
                                                                              & \multirow{-2}{*}{0.6595} & \multicolumn{2}{c|}{using Krivine~\cite{Kri79}}                 & \multirow{-2}{*}{$\infty$} & 1979                   \\\cline{2-6}
    \parbox[t]{5mm}{\multirow{-5}{*}{\rotatebox[origin=c]{90}{Lower bounds}}} & 0.5                      & \multicolumn{2}{c|}{Werner~\cite{Wer89}}                        & $\infty$                   & 1989                   \\\hline
  \end{tabular}
  \caption{
    Successive refinements of the bounds on $v_c^\mathrm{Wer}$, the nonlocality threshold of the two-qubit Werner states under projective measurements.
    Using $m$ measurements to simulate all projective ones is denoted by $m\sim\infty$.
  }
  \label{tab:history}
\end{table}

In this work, we tackle the general membership problem for the local polytope via methods from the field of constrained convex optimisation, where this is known as the approximate Carathéodory problem~\cite{MLVW17,CP23}.
Specifically, we rephrase the distance algorithm previously credited to Gilbert as the original Frank-Wolfe algorithm~\cite{FW56,LP66} (see~\cite{BCC+22,BRZ21} for recent reviews) to leverage the improvements brought to this algorithm over the last decade.
Combined with refinements of the proof in~\cite{HQV+17}, this allows us to improve on the bounds for the nonlocality threshold of the two-qubit Werner states under projective measurements, see \cref{tab:history}.
Furthermore, these results enable us to improve on the known bounds on the Grothendieck constant of order three.
Finally, we demonstrate the generality of the method by investigating multipartite scenarios: we establish new bounds for the nonlocality of the tripartite GHZ and W states~\cite{GHZ89,DVC00}, giving evidence that the latter has a nonlocality threshold under projective measurements strictly higher than the former.

\textit{Preliminaries.---}
Consider a bipartite scenario in which two parties, Alice and Bob, upon receiving inputs $x$ and $y$ chosen in $\{1\ldots m\}$, provide outputs $a$ and $b$ being $\pm1$, respectively.
Here, we are only interested in the correlation matrix arising from this process, namely, the $m\times m$ real matrix whose $(x,y)$-entries are the expectation values of $\langle ab\rangle$ with inputs $x$ and $y$.
Our specific choice of setup indeed makes marginals, i.e., expectation values $\langle a\rangle$ and $\langle b\rangle$, irrelevant as they always vanish.
Note that this will no longer be the case in multipartite scenarios, as discussed below.
In the following, correlation matrices are always denoted with bold symbols.

Classical correlation matrices all lie in the convex hull of the so-called deterministic strategies, which are rank-one matrices $\bd_{\va,\vb}$ with entries $a_xb_y$, where ${\va=(a_1,\ldots,a_m)}$ and $\vb=(b_1,\ldots,b_m)$ have $\pm1$ components.
Since ${\bd_{-\va,-\vb}=\bd_{\va,\vb}}$, there are $2^{2m-1}$ distinct deterministic strategies, which define the local correlation polytope $\mL_m$~\cite{BCP+14}.
We sometimes write $\lambda=(\va,\vb)$ for conciseness.

Given a shared quantum state $\rho$, i.e., a positive semidefinite Hermitian matrix with trace one, and traceless dichotomic observables $A_x$ and $B_y$, i.e., Hermitian matrices of trace zero and squaring to the identity, one can construct a correlation matrix by letting Alice and Bob measure their half of the shared state with their observables.
Formally, by the Born rule, the resulting matrix then has $(x,y)$-entries of the form $\Tr[(A_x\otimes B_y)\rho]$~\cite{NC11}.

The central problem we consider in this work is the membership problem for the local polytope $\mL_m$, which is twofold.
On the one hand, given a correlation matrix inside $\mL_m$, we seek to decompose it in terms of deterministic strategies, that is, to find a local model.
On the other hand, given a (quantum) correlation matrix outside $\mL_m$, we want to produce an explicit separating hyperplane to witness its nonlocality, that is, a Bell inequality.
We address this problem to compute bounds on the nonlocality threshold defined below.

In the bipartite case we focus on a family of two-qubit states named after Werner~\cite{Wer89}:
\begin{equation}
  \rho_v^\mathrm{Wer}=v\;\ketbra{\psi^-}+(1-v)\frac{\id}{4},
  \label{eqn:werner}
\end{equation}
where $\ket{\psi^-}=(\ket{01}-\ket{10})/\sqrt{2}$ is the two-qubit antisymmetric (or singlet) state.
Fixing the so-called visibility $v$ in \cref{eqn:werner} and applying qubit observables of the form $A_x=\va_x\cdot\vec{\sigma}$ and $B_y=\vb_y\cdot\vec{\sigma}$, where $\va_x$ and $\vb_y$ are real vectors on the unit 2-sphere (Bloch vectors) and $\vec{\sigma}=(\sigma_X,\sigma_Y,\sigma_Z)$ contains Pauli matrices, yields a correlation matrix whose $(x,y)$-entries are $-v\;\va_x\cdot\vb_y$.
When the number of inputs $m$ goes to infinity, we denote the different inputs directly by this Bloch vector: the observables are then $A_{\hat{x}}=\hat{x}\cdot\vec{\sigma}$ and $B_{\hat{y}}=\hat{y}\cdot\vec{\sigma}$, where the hat emphasises the infinite scenario.

\begin{question}
  \label{q:vc}
  With Werner states in \cref{eqn:werner}, which critical visibility $v_c^\mathrm{Wer}$ is the threshold between a local behaviour and a nonlocal one under projective measurements?
\end{question}

\textit{Previous works.---}
\cref{q:vc} has gained attention after the publication of~\cite{AGT06}, where it is linked with the computation of a mathematical constant (see below).
Increasingly more accurate bounds have been obtained since then, as outlined in \cref{tab:history}.

On the one hand, to obtain an upper bound it is sufficient to consider a scenario with a finite number of measurements $m$ and to exhibit (i) a matrix $M$ such that $\Tr(M\bd_{\va,\vb})\leq1$ for all deterministic strategies $\bd_{\va,\vb}$ ($M$~parametrises a normalised correlator Bell inequality), and (ii) measurements to be applied on both Alice's and Bob's sides and giving rise to a correlation matrix $v\bp$ violating this inequality, that is, such that $\Tr(M\bp)>1$.
If we consider $v>1/\Tr(M\bp)$, we have $\Tr(Mv\bp)>1$ so that the Bell inequality parametrised by $M$ is violated by $v\bp$.
Hence $v_c^\mathrm{Wer}\leq1/\Tr(M\bp)$.

On the other hand, methods to provide a lower bound cannot be as direct, since a membership proof is required for the \emph{infinite} scenario with all projective measurements.
Refs.~\cite{CGRS16,HQV+16} give a way to go from a finite number of measurements to an infinite one.
The idea is to simulate, up to an approximation factor, the infinite set of all measurements by means of a finite number of them, then to algorithmically construct a local model in this finite case, and eventually to convert the membership proof obtained there to a certificate valid for all projective measurements.

More formally, in our case, the approximation amounts to choosing $m$ measurements used both by Alice and Bob, and to computing the radius $\eta$ of the largest sphere that fits in the polyhedron defined by the vertices $\va_x$ and $-\va_x$ (adding these vectors is necessary as traceless dichotomic qubit observables correspond to two antipodal points on the Bloch sphere).
Then any shrunk direction $\eta\hat{x}$ can be, by definition of $\eta$, written as a convex mixture of the vectors $\va_x$, i.e., $\eta\hat{x}=\sum_xp_x^{\hat{x}}\va_x$; similarly, $\eta\hat{y}=\sum_yq_y^{\hat{y}}\vb_y$.
Now if we can decompose the correlation matrix with entries $-v_0\;\va_x\cdot\vb_y$ in terms of deterministic strategies, then the following equality gives a decomposition for the infinite scenario with visibility $\eta^2v_0$:
\begin{equation}
  -\eta^2v_0\;\hat{x}\cdot\hat{y}=-v_0\;(\eta\hat{x})\cdot(\eta\hat{y})=\sum_{x,y}p_x^{\hat{x}}q_y^{\hat{y}}(-v_0\;\va_x\cdot\vb_y).
  \label{eqn:step}
\end{equation}

Ref.~\cite{HQV+17} uses a polyhedron with $m=625$ measurements and finds a way to make the numerical decomposition completely analytical at the expense of an analyticity factor $\nu_1$ that we discuss below.
They eventually obtain $v_c^\mathrm{Wer}\geq\eta^2\nu_1 v_0\approx0.6829$ where
\begin{equation}\nonumber
  \eta\geq\cos\left(\frac{\pi}{50}\right)^2\approx0.9961,\ \nu_1=0.999,\ \text{and}\ v_0=0.689.
\end{equation}
In this work, we improve on all three factors, for each of them with a different theoretical reason: we first explain how to obtain polyhedra with a better shrinking factor $\eta$ (for a fixed number of measurements), then argue that our algorithm makes it possible to choose an initial visibility $v_0$ closer to the critical one, and refine the last step to have $\nu_2>\nu_1$ closer to 1.

\textit{Choosing the measurements.---}
The first step is to select the $m$ measurements that Alice and Bob perform.
The more measurements we consider, the better the approximation of the set of all projective measurements is ($\eta$ increases as the corresponding polyhedron approaches the sphere).
However, the optimisation problem on the resulting correlation polytope is also more difficult to solve, since the dimension of the corresponding space grows quadratically with $m$.
Moreover, since we want the final result to be analytical, this $\eta$ should have a closed form.

In~\cite{HQV+17}, this last necessity leads the authors to introduce a family of measurements corresponding to quite regular polyhedra and whose shrinking factors $\eta$ enjoy a relatively tight analytical lower bound.
These shrinking factors, however, are not competitive compared to polyhedra with a similar number of measurements.

Here we take a different approach to improve the quality of the shrinking factor while not losing the analyticity.
For this, we start by getting symmetric polyhedra with very good shrinking factors~\footnote{\href{https://levskaya.github.io/polyhedronisme}{https://levskaya.github.io/polyhedronisme}, see the fork \href{https://github.com/sebastiendesignolle/polyhedronisme}{https://github.com/sebastiendesignolle/polyhedronisme} which displays the shrinking factor and implements the operation S projecting vertices on the sphere}.
Then we take rational approximations of these polyhedra; importantly, we ensure that the rational points are also on the unit sphere; this approximation can be arbitrarily good as the set of rational points on the unit sphere is dense in the unit sphere.
Eventually we can compute all faces analytically and hence obtain the square of the shrinking factor $\eta^2$ as a rational.
We refer to \cref{app:polyhedra} for details on this rational approximation.

\textit{Frank-Wolfe algorithms.---}
After selecting a polyhedron as outlined above, we can construct the correlation matrix $\bp$ with entries ${\bp_{x,y}=-\va_x\cdot\vb_y}$, where $\va_x$ and $\vb_y$ are pairs of antipodal points in the chosen polyhedron; this corresponds to setting $v=1$ in \cref{eqn:werner}.
In order to obtain the distance between the local polytope $\mL_m$ and a point $v_0\bp$ on the line between $\mathbf{0}$ and $\bp$, we can choose a local point $\bx_0$ and run \cref{algo:gilbert}~\cite{BNV16,MW16}.
There, $\|\by\|_2$ denotes the 2-norm of the vectorised matrix $\by$.

\begin{algorithm}[H]
  \caption{Gilbert's algorithm~\cite{Gil66}}
  \label{algo:gilbert}
  \begin{algorithmic}[1]
    \For{$t = 0\dots T-1$}
    \State $\omega_t = \argmin_\lambda\innp{\bx_t-v_0\bp, \bd_\lambda}$ \label{line:fw_lmo}
    \State $\gamma_t = \argmin_{\gamma\in[0,1]} \|\gamma\bx_t + (1-\gamma)\bd_{\omega_t}-v_0\bp\|_2^2$ \label{line:fw_step}
    \State $\bx_{t+1} = \gamma_t \bx_t + (1-\gamma_t)\bd_{\omega_t}$
    \EndFor
  \end{algorithmic}
\end{algorithm}

As the number of deterministic strategies $\bd_\lambda$ to explore in \cref{line:fw_lmo} is exponential in $m$ (here, $2^{2m-1}$) a heuristic approach is performed, similarly to Refs.~\cite{BNV16,MW16,HQV+17}; we refer to \cref{app:qubo}.
Note that \cref{algo:gilbert} can only supply a reliable membership proof when $v_0\bp$ belongs to the local polytope.
In this case, the decomposition that the algorithm produces is valid regardless of the potential suboptimality due to the heuristic.

Although Ref.~\cite{BNV16} credits Gilbert~\cite{Gil66} for this algorithm, this instance coincides with the original Frank-Wolfe algorithm~\cite{FW56,LP66} where the function to minimise is $f(\bx)=\frac12\|\bx-v_0\bp\|_2^2$ so that $\nabla f(\bx)=\bx-v_0\bp$.
This projection-free first-order algorithm has seen a regained interest in the last decade~\cite{Jag13}, and several lines of improvement have been proposed to upgrade its convergence in various settings.
Of particular importance is the mitigation of the so-called zig-zagging behaviour~\cite{LJ15}: when optimising over polytopes, if the optimum is located near or on a facet, the original algorithm alternately selects the vertices defining this facet and moves towards them.
This leads to ever-smaller improvements in the objective function value, as the gradient becomes more orthogonal to the steps performed.
This forces the authors of Ref.~\cite{HQV+17} to pick $v_0$ such that the starting point $v_0\bp$ lies sufficiently deep inside the local polytope to ensure convergence in a reasonable time.
Here we employ a refined version of this algorithm which stores a subset of the vertices of $\mL_m$ to speed up the computations~\cite{TTP21}.
Importantly it uses so-called pairwise steps in which the current iterate moves along a line between a pair of stored vertices to decrease the weight of an unfavourable one.
This has two major benefits: the zig-zagging behaviour is reduced and the resulting decomposition is sparser.
We illustrate the zig-zagging phenomenon and describe this improved algorithm in more detail in \cref{app:bpcg}.

We implemented our approach in Julia~\cite{Julia}, based on the library \texttt{FrankWolfe.jl}~\cite{BCP21}.
Our code is freely accessible as a Julia library entitled \texttt{BellPolytopes.jl}~\footnote{\href{https://github.com/ZIB-IOL/BellPolytopes.jl}{https://github.com/ZIB-IOL/BellPolytopes.jl}}.
Importantly, it is not restricted to the case considered above but can tackle scenarios with any numbers of parties and inputs (identical for all parties), but only two outputs.
We give a few results on the multipartite case below.

\textit{Analytical decomposition.---}
After choosing an initial visibility $v_0$, we run our algorithm until the last iterate $\bx_T$ satisfies $\|\bx_T-v_0\bp\|_2\leq\epsilon$ for a chosen precision $\epsilon$.
Since we need an \emph{exact} decomposition of a point $v\bp$ to be able to certify that $v\leq v_c^\mathrm{Wer}$, such a numerical proximity is, however, insufficient.
Ref.~\cite{HQV+17} gives the following solution to overcome this difficulty: fix a factor $\nu_1$ close to 1, write $\nu_1 v_0\bp=\nu_1 \bx_T+(1-\nu_1)\by$ by suitably defining $\by$, and exhibit a local decomposition for $\by$ by using the fact that its entries are, by construction, small.

More precisely, Ref.~\cite{HQV+17} shows that any point $\by$ such that $\|\by\|_1\leq1$ has a local model.
In \cref{app:ana} we tighten this result and demonstrate that this even holds for ${\|\by\|_2\leq1}$.
This seemingly small change has in fact practical consequences as it is less restrictive on the quality of the algorithm output: for instance, the result from~\cite{HQV+17} immediately jumps from $0.6829$ to $0.6836$.
Our factor $\nu_2$ then reads
\begin{equation}
  \nu_2=\frac{1}{1+\|\bx_T-v_0\bp\|_2};
  \label{eqn:nu}
\end{equation}
we refer to \cref{app:ana} for details.

\textit{Lower bound.---}
Having outlined all steps of the proof, we can now give the computational settings that we use to attain our lower bound on $v_c^\mathrm{Wer}$.
We choose $m=406$ measurements yielding a shrinking factor of ${\eta\approx0.9968}$.
We tested several initial visibilities $v_0$ and selected ${v_0=0.692}$; for this value, the objective function steadily decreases.
Using our algorithm, we end up with 78747 deterministic strategies (out of all $2^{811}$ of them) reproducing the correlation matrix $v_0\bp$ up to $\epsilon=2\times10^{-4}$, corresponding to a factor $\nu_2\approx0.9998$ in \cref{eqn:nu}.
Finally, we recompute the decomposition with rational weights (as in~\cite{HQV+17}) to get an analytical expression of $\nu_2$.
Combining all the steps, we obtain the final analytical lower bound
\begin{equation}
  v_c^\mathrm{Wer}\geq v_\mathrm{low}=\eta^2\nu_2 v_0\approx0.6875,
  \label{eqn:vlow}
\end{equation}
whose analytical value is given in a Julia file accompanying this article.

The computation has been performed on a 64-core Intel$^\circledR$ Xeon$^\circledR$ Gold 6338 machine with \SI{512}{\giga\byte} of RAM and took about a month.
Note that this long runtime is due to the fact that we want to test our methodology extensively; this alone is not responsible for the improvement over~\cite{HQV+17}.
Owing to our theoretical improvements, we can indeed reproduce the bound therein in about 20 hours and with only 181 measurements.

\textit{Upper bound.---}
One can extract a separating hyperplane from the result of Frank-Wolfe algorithms, specifically, by taking the gradient at an approximately optimal solution.
This property is already used in Refs.~\cite{BNV16,DBV17} to construct Bell inequalities with a high resistance to noise.
The difficulty to improve on these works lies in the computation of the local value of the Bell inequality provided by the algorithm~\cite{AII06,AI07,AMO09}.
Interestingly, however, this problem can be converted into a Quadratic Unconstrained Binary Optimisation (QUBO) instance, a class of problems which has seen some recent improvements, see~\cite{RKS22} and references therein.

We had access to a version of the solver from~\cite{RKS22}.
With an initial set of 97 measurements, we ran our algorithm starting from $v_0=0.6964$ and fed the QUBO solver with the resulting hyperplane in order to obtain, in about half an hour, the local bound
\begin{equation}
  v_c^\mathrm{Wer}\leq v_\mathrm{up}\approx0.6961,
  \label{eqn:vup}
\end{equation}
whose analytical value can be found in the supplemental file together with the corresponding Bell inequality.
We refer to \cref{app:qubo} for the formulation of the local bound computation as a QUBO.
Importantly, this bound is also analytical as the Bell inequality used has integer entries, so that the decisions made in the QUBO solver that we use~\cite{RKS22} are exact.

\textit{Observations.---}
Our bounds in \cref{eqn:vlow,eqn:vup} have two immediate consequences, already described in~\cite{HQV+17} and which we only summarise here.

First, the procedure described above to construct local models for all projective measurements can be extended to general (not necessarily projective) measurements.
This is possible because qubit measurements can be simulated by projective ones up to a factor of $\sqrt{2/3}$, see~\cite[Lemma~2]{HQV+17} or~\cite{OGWA17}.
Therefore, \cref{eqn:vlow} for projective measurements gives the following lower bound on the nonlocality threshold $v_\mathrm{POVM}^\mathrm{Wer}$ for two-qubit Werner states under positive operator-valued measures (POVM):
\begin{equation}
  v_\mathrm{POVM}^\mathrm{Wer}\geq\frac23 v_c^\mathrm{Wer}\geq \frac23 v_\mathrm{low}\approx0.4583.
  \label{eqn:POVM}
\end{equation}

Second, there is a formal correspondence between the construction of local hidden variable models for two-qubit Werner states and the Grothendieck constant of order three $K_G(3)$.
In a way, this constant quantifies the power of higher dimensions for a specific task.
More precisely, if we are given any square matrix $M=[m_{xy}]$ such that $|\sum_{x,y}m_{xy}a_xb_y|\leq1$ for all scalars $a_x$ and $b_y$ in the one-dimensional unit sphere (i.e., the segment $[-1,1]$), then the Grothendieck constant of order $n$, denoted $K_G(n)$, is such that $|\sum_{x,y}m_{xy}\va_x\cdot\vb_y|\leq K_G(n)$ for all vectors $\va_x$ and $\vb_y$ in the $n$-dimensional unit sphere.
While $K_G(2)=\sqrt2$ is known~\cite{Kri79}, this is not the case of $K_G(3)$, which turns out to be precisely the inverse of $v_c^\mathrm{Wer}$~\cite{AGT06}.
Our results in \cref{eqn:vlow,eqn:vup} then directly translate into the following bounds:
\begin{equation}
  1.4367\approx\frac{1}{v_\mathrm{up}}\leq K_G(3)\leq\frac{1}{v_\mathrm{low}}\approx1.4546,
  \label{eqn:KG3}
\end{equation}
whose analytical values can be found in the supplemental file.

Other applications directly benefit from the improvement of the bounds on $v_c^\mathrm{Wer}$, such as quantum key distribution~\cite{FBL+21} or prepare-and-measure scenarios~\cite{DMBV22}.

\textit{Multipartite case.---}
The entire procedure naturally generalises to multipartite scenarios.
One important difference, however, is that marginals no longer vanish; hence, we must take them into account and reproduce them in the local model.
Computationally, it is also harder to compute a good direction in the larger correlation space, hence we are restricted to a smaller number of measurements.
We summarise our results in the tripartite case in \cref{tab:tripartite}; to the best of our knowledge, lower bounds comparable to ours are unprecedented.
We refer to \cref{app:multipartite} for details.

\begin{table*}[ht]
  \begin{tabular}{|c|c|c|c|c|}
    \hline
                                                                       & $v_c^{\mathrm{GHZ}_3}$ & Reference                                     & \#Inputs       & ~Year~                 \\\hline
                                                                       & 0.5                    & GHZ~\cite{GHZ89}                              & $2$            & 1989                   \\\cline{2-5}
                                                                       & ~0.4961~               & V\'ertesi and P\'al~\cite{VP11}               & $5$            & 2011                   \\\cline{2-5}
                                                                       & 0.4932                 & Brierley et al.~\cite{BNV16}                  & $16$           & 2016                   \\\cline{2-5}
    \parbox[t]{5mm}{\multirow{-4}{*}{\rotatebox[origin=c]{90}{Upper}}} & 0.4916                 &                                               & $16$           &                        \\\cline{1-2}\cline{4-4}
                                                                       & 0.4637                 & \multirow{-2}{*}{This work}                   & $61\sim\infty$ & \multirow{-2}{*}{2023} \\\cline{2-5}
                                                                       & 0.232                  & ~Cavalcanti et al.~\cite{CGRS16}~             & $12\sim\infty$ & 2016                   \\\cline{2-5}
                                                                       &                        &                                               & ~Entanglement~ &                        \\
    \parbox[t]{5mm}{\multirow{-4}{*}{\rotatebox[origin=c]{90}{Lower}}} & \multirow{-2}{*}{0.2}  & \multirow{-2}{*}{D\"ur and Cirac~\cite{DC00}} & threshold      & \multirow{-2}{*}{2000} \\\hline
  \end{tabular}
  \hspace{0.4cm}
  \begin{tabular}{|c|c|c|c|c|}
    \hline
                                                                       & $v_c^{\mathrm{W}_3}$     & Reference                                    & \#Inputs       & ~Year~                 \\\hline
                                                                       & ~0.6442~                 & Sen(De) et al.~\cite{SSW+03}                 & $2$            & 2003                   \\\cline{2-5}
                                                                       & 0.6007                   & Gruca et al.~\cite{GLZ+10}                   & $5$            & 2010                   \\\cline{2-5}
                                                                       & 0.5956                   & Pandit et al.~\cite{PBM+22}                  & $6$            & 2022                   \\\cline{2-5}
    \parbox[t]{5mm}{\multirow{-4}{*}{\rotatebox[origin=c]{90}{Upper}}} & 0.5482                   &                                              & $16$           &                        \\\cline{1-2}\cline{4-4}
                                                                       & 0.4911                   & \multirow{-2}{*}{This work}                  & $61\sim\infty$ & \multirow{-2}{*}{2023} \\\cline{2-5}
                                                                       & 0.228                    & ~Cavalcanti et al.~\cite{CGRS16}~            & $12\sim\infty$ & 2016                   \\\cline{2-5}
                                                                       &                          &                                              & ~Entanglement~ &                        \\
    \parbox[t]{5mm}{\multirow{-4}{*}{\rotatebox[origin=c]{90}{Lower}}} & \multirow{-2}{*}{0.1780} & \multirow{-2}{*}{Chen and Jiang~\cite{CJ20}} & threshold      & \multirow{-2}{*}{2011} \\\hline
  \end{tabular}
  \caption{
    Summarised history of the successive refinements of the bounds on the nonlocality threshold for three-qubit GHZ (left) and W (right) states under projective measurements.
    Importantly, although the 16 measurements used for the GHZ state are exactly the same as in~\cite{BNV16} (a regular polygon on the XY plane), we can reach a more robust Bell inequality owing to the improved algorithm we are using.
    The 16 measurements used for the W state correspond to a pentakis dodecahedron.
    Remarkably, $v_c^{\mathrm{GHZ}_3}<v_c^{\mathrm{W}_3}$ arises as a consequence of our bounds.
  }
  \vspace{-10pt}
  \label{tab:tripartite}
\end{table*}

\textit{Conclusion.---}
In this work, we construct local models and Bell inequalities by using Frank-Wolfe algorithms in local polytopes with binary outcomes and arbitrarily many inputs and parties.
Our main application is to improve the bounds on the nonlocality threshold of the two-qubit Werner states, hence on the Grothendieck constant of order three.
We also investigate multipartite states and find new bounds for GHZ and W states, far above their entanglement thresholds.
This opens a practical way to a better understanding of the nonlocality properties of these states.
To facilitate the reuse of the tools that we developed, we provide a Julia library with our implementation~\cite{Note2}.

A natural extension would be to increase the number of outcomes of the scenario, the algorithm working exactly the same way in the probability space.
In the qubit case, the range concerning the nonlocality threshold of two-qubit Werner states under general (not necessarily projective) measurements remains indeed wide open and a good approximation of the set of general measurements may help reduce this gap.
In higher dimensions, this extension would also require a suitable approximation of the set of projective measurements, a difficulty that was already mentioned in~\cite{HQV+16}.
Following our approach here to construct good polyhedra in the Bloch sphere, we expect symmetric measurements such as those in~\cite{NDBG20} to provide good seeds for the exploration of this direction.

More generally, the progress made in the constrained convex optimisation community and leveraged in this work could benefit all existing applications of Frank-Wolfe algorithms, e.g., for entanglement detection~\cite{SG18,Wie22}, and could also help find new utilisations, for instance, for large-scale semidefinite programming problems.

\textit{Acknowledgements.---}
The authors are grateful to Flavien Hirsch for his assistance in drawing the connection between Frank-Wolfe algorithms and Gilbert's algorithm, to Daniel Rehfeldt for allowing us to utilise his QUBO solver~\cite{RKS22}, to Tam\'as V\'ertesi for providing us with useful references and with data we used to verify our algorithm, and to Mateus Araújo, Nicolas Brunner, M\'at\'e Farkas, Antonio Frangioni, and Leo Liberti for their contribution in the discussions.
This research was partially funded by the DFG Cluster of Excellence MATH+ (EXC-2046/1, project id 390685689) funded by the Deutsche Forschungsgemeinschaft (DFG).

\textit{Erratum.---}
GPT 6 Astra was used to detect and fix the mistake in \cref{lem:ball} with marginals, as well as to correct minor grammatical and mathematical errors.
Moreover, marginals were not handled properly and multipartite computations had to be run again.
The strict separation between the robustnesses of the tripartite GHZ and W states did not survive the corresponding modifications.
Increasing the number of measurements would undoubtedly allow us to repair this proof, but this would go beyond the scope of the initial article.

\bibliography{DIB+23}

\appendix
\onecolumngrid

\newpage

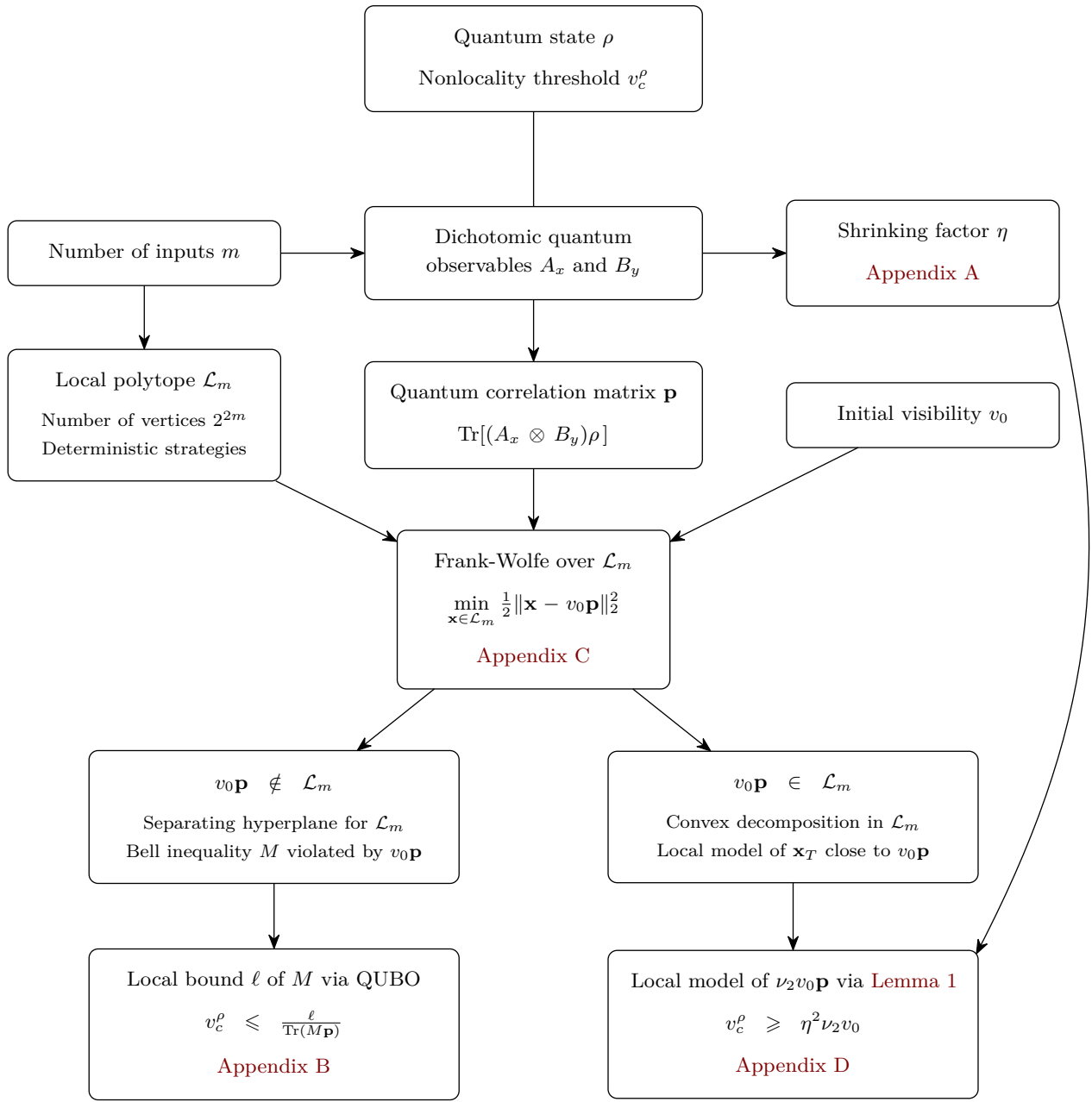
\begin{figure}[H]
  \begin{tikzpicture}[every node/.style={inner sep=10, outer sep=0}]
    \node (rho) [block, text width=4.5cm]{
      Quantum state $\rho$\\[8pt]
      Nonlocality threshold $v_c^\rho$
    };
    \node (A) [block, below of=rho, text width=4.5cm]{
      Dichotomic quantum\\[2pt] observables $A_x$ and $B_y$
    };
    \node (m) [block, left of=A, xshift=-3cm, text width=3.5cm]{
      Number of inputs $m$\\[8pt]
    };
    \node (p) [block, below of=A, yshift=0.5cm, text width=4.5cm]{
      Quantum correlation matrix $\bp$\\[8pt]
      $\Tr[(A_x\otimes B_y)\rho\,]$
    };
    \node (L) [block, below of=m, yshift=0.5cm, text width=3.5cm]{
      Local polytope $\mL_m$\\[8pt]
      {\footnotesize
        Number of vertices $2^{2m-1}$\\[2pt]
        Deterministic strategies
      }
    };
    \node (eta) [block, right of=A, xshift=3cm, text width=3.5cm]{
      Shrinking factor $\eta$\\[8pt]
      \cref{app:polyhedra}
    };
    \node (v0) [block, below of=eta, yshift=0.5cm, text width=3.5cm]{Initial visibility $v_0$};
    \node (fw) [block, below of=p, text width=3.5cm]{
      Frank-Wolfe over $\mL_m$\\[8pt]
      $\min\limits_{\bx\in\mL_m}\frac{1}{2}\|\bx-v_0\bp\|_2^2$\\[8pt]
      \cref{app:bpcg}
    };
    \node (fwub) [block, below of=fw, xshift=-4cm, yshift=-0.2cm, text width=5cm]{
      $v_0\bp \notin \mL_m$\\[8pt]
      {\footnotesize
        Separating hyperplane for $\mL_m$\\[2pt]
        Bell inequality $M$ violated by $v_0\bp$
      }
    };
    \node (fwlb) [block, below of=fw, xshift=4cm, yshift=-0.2cm, text width=5cm]{
      $v_0\bp \in \mL_m$\\[8pt]
      {\footnotesize
        Convex decomposition in $\mL_m$\\[2pt]
        Local model of $\bx_T$ close to $v_0\bp$
      }
    };
    \node (ub) [block, below of=fwub, yshift=-0.2cm, text width=5cm]{
      Local bound $\ell$ of $M$ via QUBO\\[8pt]
      $v_c^\rho\leq\frac{\ell}{\Tr(M\bp)}$\\[8pt]
      \cref{app:qubo}
    };
    \node (lb) [block, below of=fwlb, yshift=-0.2cm, text width=5cm]{
      Local model of $\nu_2 v_0\bp$ via \cref{lem:ball}\\[8pt]
      $v_c^\rho\geq\eta^2 \nu_2 v_0$\\[8pt]
      \cref{app:ana}
    };
    \draw [line width=0.5pt] (rho) -- (A);
    \draw [line] (m) to (A);
    \draw [line] (m) to (L);
    \draw [line] (A) to (p);
    \draw [line] (A) to (eta);
    \draw [line] (v0) -- (fw);
    \draw [line] (p) -- (fw);
    \draw [line] (L) -- (fw);
    \draw [line] (fw) -- (fwub);
    \draw [line] (fw) -- (fwlb);
    \draw [line] (fwub) -- (ub);
    \draw [line] (fwlb) -- (lb);
    \draw [line, shorten <= -2pt, shorten >= -1.8pt] (eta.south east) to [bend left=20] (lb.north east);
    \draw [line, color=white] (m.south west) to [bend right=20] (ub.north west);
  \end{tikzpicture}
  \vspace{0.5cm}
  \caption{
    Summary of our methodology.
    The main text mostly considers the case where the shared quantum state $\rho$ is the two-qubit singlet state, see \cref{eqn:werner,q:vc}.
    Given a number of inputs $m$, the local polytope $\mL_m\subseteq\mathbb{R}^{(m+1)^2-1}$ is the convex hull of all $2^{2m}$ vertices $\bd_\lambda$, where the different $\lambda$ account for all deterministic strategies assigning the inputs on Alice's and Bob's sides to $\pm1$.
    For two-qubit shared states, the observables describing a quantum strategy are encoded as $A_x = \va_x \cdot \vec{\sigma}$ and $B_y = \vb_y \cdot \vec{\sigma}$, where $\vec{\sigma}=(\sigma_X,\sigma_Y,\sigma_Z)$ contains the Pauli matrices; the $m$ Bloch vectors $\va_x$ (and $\vb_y$) are selected so that the corresponding polyhedron in the Bloch sphere has a shrinking factor $\eta$ as close to 1 as possible, that is, an inscribed sphere as large as possible.
    The initial visibility is selected to be slightly higher than the putative value of $v_c^\rho$ (estimated empirically) to obtain an upper bound (left branch of the diagram), or slightly lower to obtain a lower bound (right branch).
    For the upper bound \cref{eqn:vup} in the main text, the number of inputs $m=97$ is chosen such that finding the local bound remains computationally tractable.
    For the lower bound \cref{eqn:vlow} in the main text, the number of inputs $m=406$ is chosen such that \cref{algo:bpcg} finds in a reasonable time $\bx_T$ such that $\|\bx_T-v_0\bp\|<10^{-3}$, so that the analyticity factor $\nu_2$ in \cref{eqn:nu} is close enough to 1 to provide a good lower bound.
    Although the diagram focuses on the bipartite case, the approach seamlessly generalises to multipartite scenarios, as outlined in \cref{app:multipartite}.
  }
  \label{fig:pipeline}
\end{figure}

\newpage

\begin{figure}[ht]
  \centering
  \subfloat{\includegraphics[width=0.2\columnwidth]{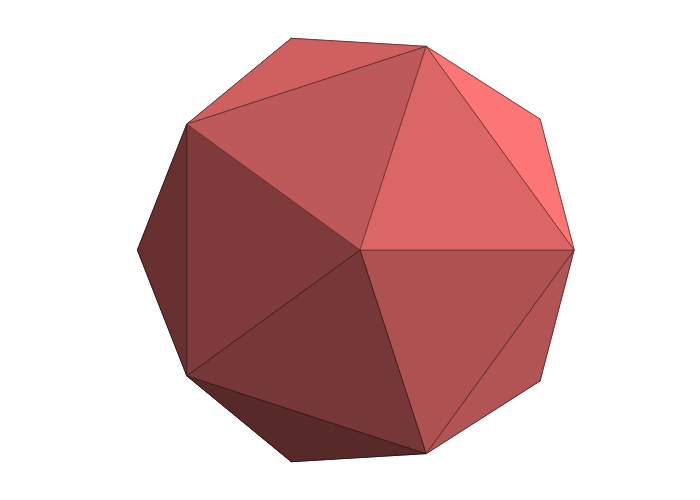}}
  \subfloat{\includegraphics[width=0.2\columnwidth]{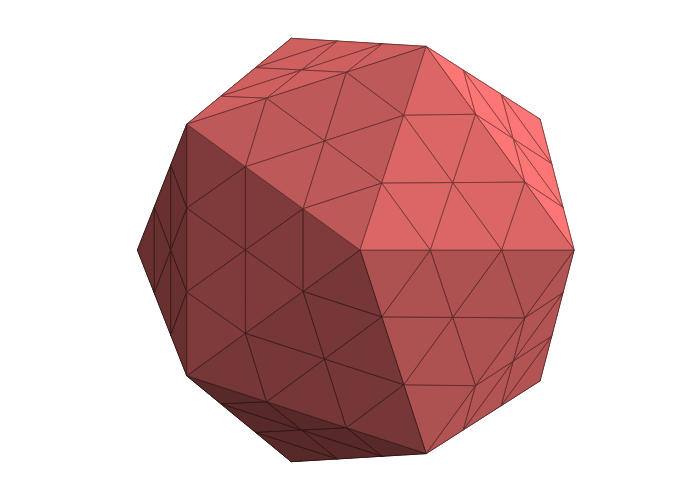}}
  \subfloat{\includegraphics[width=0.2\columnwidth]{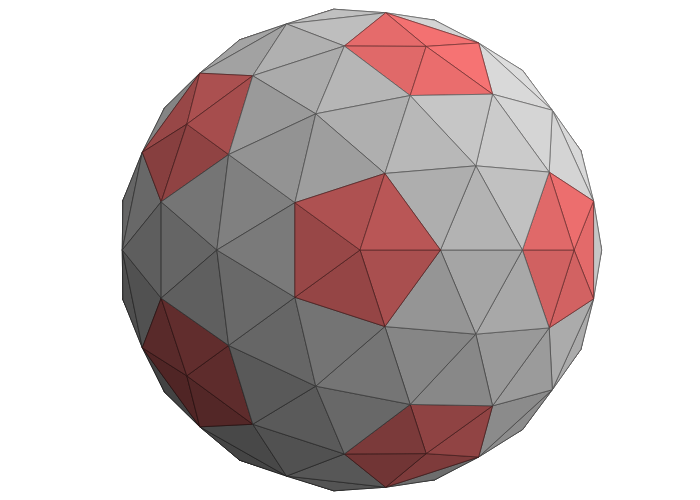}}
  \subfloat{\includegraphics[width=0.2\columnwidth]{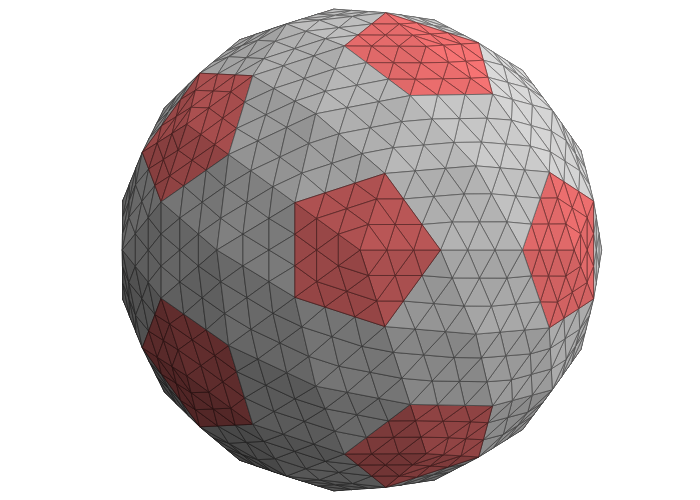}}
  \subfloat{\includegraphics[width=0.2\columnwidth]{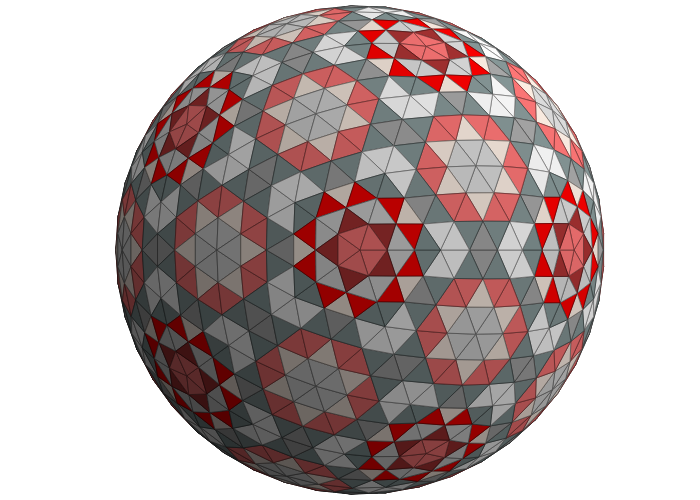}}
  \caption{
    From left to right: starting from an icosahedron inscribed in the unit sphere (12 vertices), we divide each triangular face into 9 smaller triangles, project all vertices onto the unit sphere, and repeat this procedure one more time.
    Colours only help visualise the symmetry.
    The resulting polyhedron on the right has a shrinking factor (also called inradius) of 0.9968.
    Note that directly dividing each triangular face of the icosahedron into 81 smaller triangles yields a slightly smaller shrinking factor, namely, 0.9965.
    The images were obtained using~\cite{Note1}.
  }
  \label{fig:pol}
\end{figure}

\section{Analytical polyhedra}
\label{app:polyhedra}

In this section, we present the method to generate rational polyhedra on the unit sphere with a good shrinking factor, that is, an inscribed sphere with a large radius.

First, we craft symmetric polyhedra with a good shrinking factor.
This polyhedron generation process is essentially empirical, partly because the optimisation of a distribution of points on the sphere is a notoriously hard problem, being equivalent to the so-called Thomson problem~\cite{Tho04}, which aims at placing electrons on the sphere in a way that minimises the total electrostatic energy.
Analytic solutions to this problem are only known for a small number of electrons (up to twelve with the icosahedron) so that, for our similar problem, a good enough configuration is the best we can hope for.

First, we choose a Platonic solid (typically, the icosahedron) and apply various transformations on it~\cite{Note1}, for instance, subdividing each triangular face into smaller triangles or projecting all vertices onto the unit sphere.
We give simple examples in \cref{fig:pol}.
Note that Ref.~\cite{DBV17} uses a heuristic algorithm to construct well-spread distributions; we do not know how their approach compares to the one presented here.

Then, we round up the vertices in a suitable way that maintains the resulting rational points on the unit sphere.
To this end, we take advantage of the following parametrisation of the sphere: firstly, we express a point $s\in S^2$ given in Cartesian coordinates $(x, y, z)$ by spherical coordinates $(\varphi,\theta)$ such that $(x,y,z)=(\sin\varphi\cos\theta,\sin\varphi\sin\theta,\cos\varphi)$ and, secondly, we use known trigonometric transformations to write
\begin{equation}
  (x,y,z)=\left(\frac{2t_\varphi}{1+t_\varphi^2}\cdot\frac{1-t_\theta^2}{1+t_\theta^2}\;\;,\;\frac{2t_\varphi}{1+t_\varphi^2}\cdot\frac{2t_\theta}{1+t_\theta^2}\;\;,\;\frac{1-t_\varphi^2}{1+t_\varphi^2}\right),
  \label{eqn:trick1}
\end{equation}
where we define
\begin{equation}
  t_\varphi=\tan\frac{\varphi}{2}\qquad\text{and}\qquad t_\theta=\tan\frac{\theta}{2}.
  \label{eqn:trick2}
\end{equation}
The crucial property in \cref{eqn:trick1} is that it preserves rationality, so that, by taking a rational approximation in \cref{eqn:trick2}, we eventually obtain a rational approximation $\tilde{s}\in S^2$ of the initial point $s$, also lying exactly on the unit sphere.

Finally, we can compute the faces $f$ of the polyhedron, that is, inequalities $\innp{a_f,r}\leq\beta_f$ (with unit normal vectors $a_f$) such that the point $r\in\mathbb{R}^3$ is within the polyhedron if and only if it satisfies all of them.
The shrinking factor is simply the smallest distance from the centre of the sphere to the different faces, that is, $\eta^2=\min_f\beta_f^2$.

The list of all polyhedra used throughout this article is given in \cref{tab:polyhedra}.
In practice, we use the library \texttt{Polyhedra.jl}~\cite{Polyhedra}; conveniently, since the Julia language enjoys multiple dispatch, the very same method can be called on the initial polyhedron (with elements of type \texttt{Float64}) or its rational approximation (with elements of type \texttt{Rational\{BigInt\}}).

Notice that, although the bounds provided in the text file accompanying this article use this general method to have a rational correlation and thus be able to perform all final computations over the rationals, some shrinking factors of the polyhedra in \cref{tab:polyhedra} can be directly computed.
Here we provide these irrational values for completeness:
\begin{equation}\nonumber
  \eta_6^2=\frac{5+2\sqrt5}{15},\ \eta_{16}^2=\frac{620+185 \sqrt{5}+\sqrt{30 \left(12905+5701 \sqrt{5}\right)}}{2245},\ \eta_{46}^2=\frac{3 \left(2470+63 \sqrt{5}+\sqrt{30 \left(110429+39255 \sqrt{5}\right)}\right)}{16045},
\end{equation}
and $\eta_{406}^2$ is the largest root of a polynomial of degree eight given in the supplemental file.

\begin{table}[ht]
  \centering
  \begin{tabular}{|c|c|c|c|c|}
    \hline
    ~~$m=\frac{\#V}{2}$~~ & ~~Shrinking factor $\eta_m$~~ & Comments                                                                       & Recipe~\cite{Note1} \\\hline
    6                     & 0.7947                        & Icosahedron, \cref{fig:pol} (left)                                             & SI                  \\\hline
    16                    & 0.9226                        & ~~Upper bound on the W state in \cref{tab:tripartite}, pentakis dodecahedron~~ & SkD                 \\\hline
    46                    & 0.9716                        & \cref{fig:pol} (middle)                                                        & Su3I                \\\hline
    61                    & 0.9792                        & Lower bounds in \cref{tab:tripartite}                                          & SASuSkD             \\\hline
    97                    & 0.9858                        & Upper bound in \cref{eqn:vup}                                                  & ~~SA2SuSAukSC~~     \\\hline
    181                   & 0.9929                        & Improves on~\cite{HQV+17} with $v_0=0.693$                                     & SdStSuSkD           \\\hline
    406                   & 0.9968                        & Lower bound in \cref{eqn:vlow}, \cref{fig:pol} (right)                         & Su3Su3I             \\\hline
  \end{tabular}
  \caption{
    List of symmetric polyhedra mentioned in this work.
    They are geodesic polyhedra (i.e., having triangular faces) with icosahedral symmetry (except for $m=97$).
    Note that the number $m$ of measurements corresponds to half of the number $\#V$ of vertices since each pair of antipodal vertices gives rise to one measurement.
  }
  \label{tab:polyhedra}
\end{table}

\section{Linear minimisation oracle}
\label{app:qubo}

In this section, we address the subproblem solved by the Frank-Wolfe algorithm in every iteration (see \cref{app:bpcg} for a full description of the algorithm).
Geometrically, in our case, this part amounts to choosing the vertex towards which we move to reduce the distance between the current iterate and the objective point.
We first discuss the heuristic used in the implementation, and finally reformulate it as a Quadratic Unconstrained Binary Optimisation (QUBO) problem to solve hard instances exactly.
Importantly, the explicit decompositions obtained with the heuristic approach are valid regardless of its potential suboptimality; hence this methodology is suited to compute lower bounds on the nonlocality threshold.

Both methods address the problem
\begin{equation}
  \max_{\va,\vb\in\{\pm1\}^m} \sum_{x,y\in[m]} a_x m_{xy} b_y,
  \label{eqn:miqp}
\end{equation}
where $M=[m_{xy}]$ is an $m\times m$ real matrix and where we denote $\va=a_1\ldots a_m$ (similarly for $\vb$ and for all vectors in the following) and $[m]=1\ldots m$.
From a quantum information perspective, this amounts to finding the local bound of the (correlator) Bell inequality parametrised by $M$.

\subsection{Alternating maximisation}
\label{app:heuristic}

Refs.~\cite{BNV16,MW16,HQV+17} provide a heuristic to quickly compute a good solution to \cref{eqn:miqp}.
This heuristic selects a random $\va$ and alternately maximises over $\vb$ or $\va$, until no more improvement is observed.
The key point is that each of these two maximisation steps is linear and thus admits a closed-form solution.

Specifically, we pick a random $\va^h\in\{\pm1\}^m$ and compute
\begin{equation}
  \vb^h = \argmax_{\vb\in\{\pm1\}^m} \sum_{y\in[m]} b_y \bigg(\sum_{x\in[m]} m_{xy}a^h_x\bigg),\quad\text{that is,}\quad b_y = \sign\bigg(\sum_{x\in[m]} m_{xy}a^h_x\bigg)
\end{equation}
to inject the resulting $\vb^h$ in
\begin{equation}
  \va^h = \argmax_{\va\in\{\pm1\}^m} \sum_{x\in[m]} a_x \bigg(\sum_{y\in[m]} m_{xy}b^h_y\bigg),\quad\text{that is,}\quad a_x = \sign\bigg(\sum_{y\in[m]} m_{xy}b^h_y\bigg).
\end{equation}
Note that, contrary to usual conventions, here $\sign(0)=1$.
We repeat the last two steps until the objective value in \cref{eqn:miqp} stops increasing.
In practical instances, this alternating maximisation algorithm does not require more than a dozen rounds to end.
Furthermore, the computed point is a valid extreme point of the polytope.

Depending on the random input, the output may achieve different values in \cref{eqn:miqp}.
To improve the quality of the result, we run it multiple times with random initialisations (a few thousand in practice) and take the best solution.
Importantly, despite having no guarantee that the resulting bound is close to the actual optimum, the heuristic can still be used in a provably convergent lazified Frank-Wolfe algorithm as a weak separation oracle~\cite{BPZ19}.

\subsection{QUBO reformulation}
\label{app:reformulation}

The problem in \cref{eqn:miqp} is NP-hard and, to the best of our knowledge, Ref.~\cite{DBV17} solves the largest generic instance, that is, such that no specific property of $M$ is exploited to speed up or bypass the computation.
The instance therein involves a matrix $M$ of size $92\times92$ and uses a custom branch-and-bound algorithm.
Here we argue that the recent developments in solving QUBO problems make the result therein obsolete~\cite{RKS22}.

Specifically, we cast \cref{eqn:miqp} as a QUBO by replacing the variables $\va,\vb$ in \cref{eqn:miqp} with new binary variables $\vu,\vv\in\{0,1\}^m$.
This yields the following formulation:
\begin{equation}
  \max_{\vu,\vv\in\{0,1\}^m} \sum_{x,y\in[m]} (2u_x - 1) m_{xy} (2v_y - 1).
  \label{eqn:qubo}
\end{equation}
The new variables $\vu,\vv$ satisfy $u_x = u_x^2$ for all $x\in[m]$ and $v_y = v_y^2$ for all $y\in[m]$; therefore, the problem in \cref{eqn:qubo} can be written as
\begin{equation}
  \max_{\vu,\vv\in\{0,1\}^m} \sum_{x,y\in[m]}\big[2m_{xy} (2u_x v_y - u_x^2 - v_y^2) + m_{xy}\big],
  \quad\text{or in a more compact form}\quad
  c-2\min_w w^\top Q w,
\end{equation}
where we introduce the constant $c=\sum_{x,y\in[m]} m_{xy}$, the variable $w=(\vu,\vv)\in\{0,1\}^{2m}$, and
\begin{equation}
  \mbox{\scalebox{1.3}{$Q=$}}
  \begin{pmatrix}
    \begin{matrix}
      \sum\limits_{y\in[m]} m_{1y} &         & \mbox{\scalebox{1.3}{$0$}}   \\
                                   & \ddots  &                              \\
      \mbox{\scalebox{1.3}{$0$}}   &         & \sum\limits_{y\in[m]} m_{my} \\[2pt]
    \end{matrix}
                                     & \rvline & \mbox{\scalebox{1.3}{$-M$}} \\\hline
    \mbox{\scalebox{1.3}{$-M^\top$}} & \rvline &
    \begin{matrix}
      \sum\limits_{x\in[m]} m_{x1} &         & \mbox{\scalebox{1.3}{$0$}}   \\
                                   & \ddots  &                              \\
      \mbox{\scalebox{1.3}{$0$}}   &         & \sum\limits_{x\in[m]} m_{xm} \\
    \end{matrix}
  \end{pmatrix}.
\end{equation}

\section{Frank-Wolfe algorithms}
\label{app:bpcg}

In this section, we give the exact variation of the original Frank-Wolfe algorithm that we use to tackle the membership problem for the local polytope.
This is based on the \emph{lazy blended pairwise conditional gradients}~\cite{TTP21}, which stores the vertices of the polytope used to represent the current iterate and reuses them to accelerate convergence.
Ideas for reusing previously discovered vertices as weak separation oracles were already suggested in \cite{BNV16} in the context of the Gilbert algorithm and \cite{BPZ19} for Frank-Wolfe.
In both contexts, this vertex storage was used as a cache only to reduce the number of calls to the true minimisation oracle and not to improve descent directions as in the present algorithm.
We recall that $\bd_{\lambda}$ is the correlation matrix associated with the deterministic strategy $\lambda=(\va,\vb)$.

\begin{figure}[ht]
  \centering
  \subfloat{\includegraphics[width=0.45\columnwidth]{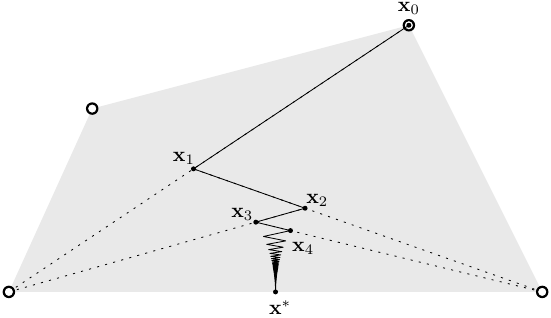}}
  \hspace{1.5cm}
  \subfloat{\includegraphics[width=0.45\columnwidth]{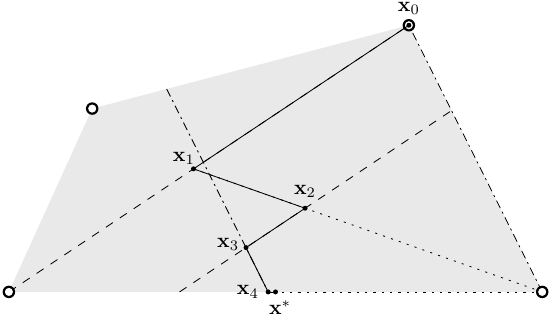}}
  \caption{
    For a quadratic function $\frac12\|\bx-\bx^*\|_2^2$, illustration of the zig-zagging effect suffered by the standard Frank-Wolfe algorithm (left), that is, \cref{algo:gilbert} in the main text, and of the countermeasure implemented by its pairwise variant (right), that is, \cref{algo:bpcg}.
    Importantly, the pairwise steps in the latter are parallel to the lines joining the worst and best cached vertices with respect to the gradient $\bx_t-\bx^*$ at the current iterate $\bx_t$.
    Note that the second pairwise step going from $\bx_3$ to $\bx_4$ (parallel to the dashed-dotted line) drops the unfavourable initial vertex.
  }
  \label{fig:zigzag}
\end{figure}

\cref{algo:bpcg} leverages an \emph{active} set representation of the current iterate, i.e., it keeps track of the convex combination of vertices (referred to as \emph{atoms} in the Frank-Wolfe literature) forming the iterate.
The \emph{pairwise} denomination comes from steps that are transferring weight in the convex combination from one \emph{away} atom (which has a nonzero weight) to another \emph{forward} one (which may or may not have a nonzero weight), as illustrated in \cref{fig:zigzag}.
Geometrically, such steps yield a descent direction parallel to the line joining the away and the forward vertices.
The \emph{lazy} aspect of the algorithm refers to the fact that the LMO from \cref{line:bpcg_lmo} is called only when the pairwise steps on the active set atoms cannot provide sufficient progress.
Since we optimise a smooth and strongly convex function over a polytope, the algorithm enjoys a linear convergence rate~\cite[Theorem~3.6(A)]{TTP21}, as long as the alternating maximisation used within the LMO (cf.~\cref{app:heuristic}) provides sufficient progress.
The computed solution is ensured to be local, and the validity of the separating hyperplane is proven with the last LMO call performed exactly via a QUBO, see \cref{app:reformulation}.

In our case, the gradient $\nabla f(\bx)$ is equal to $\bx-v_0\bp$.
We can exploit this structure to accelerate the algorithm by avoiding the repeated computation of $\innp{\nabla f(\bx_t), \bd_\lambda}$ in \cref{line:bpcg_away,line:bpcg_local} of \cref{algo:bpcg}.
More precisely, since the current iterate is a convex combination of the atoms of the active set, i.e., $\bx_t=\sum_\mu q_\mu^{(t)}\bd_\mu$, we have
\begin{equation}
  \innp{\nabla f(\bx_t), \bd_\lambda}=\sum_\mu q_\mu^{(t)}\innp{\bd_\mu,\bd_\lambda}-\innp{v_0\bp,\bd_\lambda}.
  \label{eqn:argminmax}
\end{equation}
We can then store all terms $\innp{\bd_\mu,\bd_\lambda}$, all terms $\innp{v_0\bp,\bd_\lambda}$, and maintain an up-to-date value of the sum in \cref{eqn:argminmax} according to potential changes from previous iterations.
Since at most two weights $q_\mu^{(t)}$ can have changed compared to the previous weights $q_\mu^{(t-1)}$, the cost of this update is small compared to recomputing all terms.

\begin{algorithm}[H]
  \caption{Lazy blended pairwise conditional gradients~\cite{TTP21} for the membership problem for the local polytope}
  \label{algo:bpcg}
  \begin{algorithmic}[1]
    \State Parameters: initial strategy $\lambda_0$, lazy tolerance $K \geq 1$
    \State $\mathcal{S}_0 = \{\bd_{\lambda_0}\}$, $\bx_0 = \bd_{\lambda_0}$ \Comment{initialise the active set}
    \State $\Phi_0 = f (\bd_{\lambda_0})$ \Comment{initial upper bound on the primal gap, since $f(\bx)=\frac12\|x-v_0\bp\|_2^2\geq 0$}
    \For{$t = 0\dots T-1$}
    \State $\alpha_t = \argmax_{\lambda\in \mathcal{S}_t}\innp{\nabla f(\bx_t), \bd_\lambda}$ \Comment{worst strategy from the active set, i.e., least aligned with the gradient} \label{line:bpcg_away}
    \State $\lambda_t = \argmin_{\lambda\in \mathcal{S}_t}\innp{\nabla f(\bx_t), \bd_\lambda}$ \Comment{best strategy from the active set, i.e., most aligned with the gradient} \label{line:bpcg_local}
    \If{$\innp{\nabla f(\bx_t),\bd_{\alpha_t}-\bd_{\lambda_t}}\geq\Phi_t$}
    \State $\ba_t = \bd_{\alpha_t}-\bd_{\lambda_t}$
    \State $\Gamma_t = c[\bx_t](\alpha_t)$ \Comment{coefficient of $\alpha_t$ in $\bx_t$}
    \State $\gamma_t = \argmin_{\gamma\in[0,\Gamma_t]} f(\bx_t-\gamma\ba_t)$\Comment{closed-form step size from the quadratic objective}
    \State $\Phi_{t+1} = \Phi_{t}$
    \If{$\gamma_t<\Gamma_t$}
    \State $\mathcal{S}_{t+1} = \mathcal{S}_t$ \Comment{descent step: keep the current active set}
    \Else
    \State $\mathcal{S}_{t+1} = \mathcal{S}_t\setminus\{\alpha_t\}$ \Comment{drop step: remove $\alpha_t$ from the active set}
    \EndIf
    \Else
    \State $\omega_t = \argmin_\lambda\innp{\nabla f(\bx_t), \bd_\lambda}$ \Comment{LMO: best strategy found by the heuristic in \cref{app:heuristic}} \label{line:bpcg_lmo}
    \State $\ba_t = \bx_t-\bd_{\omega_t}$
    \If{ $\innp{\nabla f(\bx_t), \bx_t - \bd_{\omega_t}} \geq \Phi_t / K$ }
    \State $\gamma_t = \argmin_{\gamma\in[0,1]} f(\bx_t-\gamma\ba_t)$
    \State $\Phi_{t+1} = \Phi_{t}$
    \State $\mathcal{S}_{t+1} = \mathcal{S}_t\cup\{\omega_t\}$ \Comment{Frank-Wolfe step: add the new atom found by the LMO}
    \Else
    \State $\gamma_t = 0$
    \State $\Phi_{t+1} = \Phi_{t} / 2$
    \State $\mathcal{S}_{t+1} = \mathcal{S}_t$
    \EndIf
    \EndIf
    \State $\bx_{t+1} = \bx_{t} -\gamma_t\ba_t$ \Comment{update the weights}
    \EndFor
  \end{algorithmic}
\end{algorithm}

\section{Analytical decomposition}
\label{app:ana}

In this section, we recall the procedure from~\cite{HQV+17} to convert the output of \cref{algo:bpcg} into an analytical local model, describe the limit faced when only resorting to~\cite[Lemma~1]{HQV+17}, and illustrate the derivation of \cref{lem:ball} below in the simple case of $2\times2$ correlation matrices.

Given the objective point $v_0\bp$ and the last iterate $\bx_T$ of the algorithm, we can write, for $R$ a real number to be chosen later,
\begin{equation}
  \nu_2v_0\bp=\nu_2\bx_T+(1-\nu_2)\by\quad\text{by defining}\quad\by=\frac{\nu_2}{1-\nu_2}(v_0\bp-\bx_T)\quad\text{and}\quad\nu_2=\frac{R}{R+\|\bx_T-v_0\bp\|_2}.
  \label{eqn:y}
\end{equation}
Note that, in case $1-\nu_2=0$, i.e., $\|\bx_T-v_0\bp\|_2=0$, such a procedure is not needed as the output of the algorithm is already a valid analytical decomposition.
Then, by construction, $\by$ satisfies $\|\by\|_2=R$ so that we can use the following result, whose proof can be found in \cref{app:lemma}, to choose $R=1$ and claim that $\by$ is local.

\begin{lemma}
  In a Bell scenario with two outputs per party, the closed unit ball for the 2-norm is contained in the local polytope of full correlators.
  \label{lem:ball}
\end{lemma}

In summary, we have exhibited a convex decomposition of $\nu_2v_0\bp$ in which $\bx_T$ has a local model constructed by the algorithm and $\by$ is also local because its 2-norm is at most 1.
Note that, as in~\cite{HQV+17}, we round up the weights of the output of the algorithm to get an analytical $\bx_T$.

\subsection{Necessity of \texorpdfstring{\cref{lem:ball}}{Lemma 1}}

Here we briefly describe the first attempts we made to obtain an analyticity factor close to 1 when resorting to~\cite[Lemma~1]{HQV+17} instead of \cref{lem:ball}, which amounts to having $\|\bx_T-v_0\bp\|_1$ instead of $\|\bx_T-v_0\bp\|_2$ in \cref{eqn:y}.
These attempts were partly successful as we could use them to improve on the lower bound from~\cite{HQV+17}.

Given the atoms describing the last iterate $\bx_T$ returned by \cref{algo:bpcg} and considering that $v_c^\mathrm{Wer}\geq\eta^2\nu_1 v_0$ where $\nu_1=1/(1+\|\bx_T-v_0\bp\|_1)$, one can try to adjust the weights of the convex decomposition to improve on the bound provided by these atoms.
This amounts to solving the following linear-fractional programming problem:
\begin{equation}
  \max_{v,\vq} \frac{v}{1+\big\|\sum_\mu q_\mu\ba_\mu-v\bp\big\|_1}\quad
  \text{such that}\quad
  \sum_\mu q_\mu\ba_\mu\geq v\bp,\quad
  v\leq1,\quad
  \sum_\mu q_\mu\leq1,\quad
  \text{and}\quad
  \forall \mu:\quad q_\mu\geq0,
  \label{eqn:frac_norm1}
\end{equation}
where $\mu$ indexes the atoms in the active set produced by \cref{algo:bpcg}.
Each atom $\ba_\mu$ is a matrix of the same dimension as $\bp$ and the constraint $\sum_\mu q_\mu\ba_\mu\geq v\bp$ is entry-wise.
Note that, with this constraint, \cref{eqn:frac_norm1} can be reformulated as a linear program~\cite{Baj03}.

The above procedure is further motivated by the relative sparsity of the decomposition provided by \cref{algo:bpcg} compared to previous methods.
For $m=181$ and $v_0=0.693$, it is indeed able to increase the analyticity factor from $\nu_1\approx0.81$ at the output of \cref{algo:bpcg} (with 18043 atoms) up to $\nu_1\approx1$, at the expense of a slightly lower $v^\ast\approx0.692996$, resulting in a lower bound $v_c^\mathrm{Wer}\geq\eta^2\nu_1v^\ast\approx0.6832$ improving on~\cite{HQV+17}.

However, when trying to go beyond this number of measurements, instances of practical interest are prohibitively expensive.
They typically involve tens of thousands of variables and hundreds of thousands of constraints.
In fact, even some of the best commercial mixed integer linear programming solvers cannot solve them in a reasonable time (a few days), and require very large amounts of RAM (about \SI{400}{\giga\byte}).
To have any hope of tackling the problem in \cref{eqn:frac_norm1} in large dimensions, one would have to rely on large-scale optimisation techniques such as, say, bundle methods (as \cref{eqn:frac_norm1} can be further reformulated as the maximisation of a concave polyhedral function over a simple polytope), or random or alternating projections to name a few.
When starting to explore these directions, we fortunately found the workaround formulated in \cref{lem:ball}.

\subsection{Illustration of \texorpdfstring{\cref{lem:ball}}{Lemma 1}}

Now we give a proof of \cref{lem:ball} in the case of $2\times2$ correlation matrices to make the understanding of the general proof, given in \cref{app:lemma}, easier.
This elaborates on~\cite[Lemma~1]{HQV+17} in which the same result is proven for the 1-norm.
In fact, we use the very same decomposition, but in a more explicit manner that allows us to interpret it in a refined way.
The proof for the general case can be found in \cref{app:lemma}.

We consider the $2\times2$ real matrix
\begin{equation}
  \br=\begin{pmatrix} \alpha & \beta\\ \gamma & \delta \end{pmatrix},
  \label{eqn:ijkl}
\end{equation}
which we want to decompose as a convex combination of the deterministic strategies whose matrices read
\begin{equation}
  \begin{pmatrix} a_1b_1 & a_1b_2\\ a_2b_1 & a_2b_2 \end{pmatrix},
  \label{eqn:ds22}
\end{equation}
with $a_1,a_2,b_1,b_2$ being $\pm1$.
Since $(a_1,a_2,b_1,b_2)$ and $(-a_1,-a_2,-b_1,-b_2)$ give the same matrix in \cref{eqn:ds22}, we fix $a_1=1$.
Furthermore, noting that $(a_1,a_2,b_1,b_2)$ and $(a_1,a_2,-b_1,-b_2)$ give opposite matrices, we fix $b_1=1$ for now.
We are then left with the four following matrices:
\begin{equation}
  \begin{pmatrix} + & +\\ + & + \end{pmatrix}\,,\,
  \begin{pmatrix} + & -\\ + & - \end{pmatrix}\,,\,
  \begin{pmatrix} + & +\\ - & - \end{pmatrix}\,,\,
  \begin{pmatrix} + & -\\ - & + \end{pmatrix}.
  \label{eqn:vertices22}
\end{equation}
In order to decompose the matrix $\br$ in \cref{eqn:ijkl} we first decompose the matrices from the standard basis; for instance, $\be_{12}$ is
\begin{equation}
  \begin{pmatrix} 0 & 1\\ 0 & 0 \end{pmatrix}=\frac14\left[
  \begin{pmatrix} + & +\\ + & + \end{pmatrix}\!-\!
  \begin{pmatrix} + & -\\ + & - \end{pmatrix}\!+\!
  \begin{pmatrix} + & +\\ - & - \end{pmatrix}\!-\!
  \begin{pmatrix} + & -\\ - & + \end{pmatrix}\right].
\end{equation}
Then, by regrouping the terms, we end up with the following decomposition
\begin{align}
  \br
  =\frac{\alpha+\beta+\gamma+\delta}{4}
  \begin{pmatrix} + & +\\ + & + \end{pmatrix}
  +\frac{\alpha-\beta+\gamma-\delta}{4}
  \begin{pmatrix} + & -\\ + & - \end{pmatrix}
  +\frac{\alpha+\beta-\gamma-\delta}{4}
  \begin{pmatrix} + & +\\ - & - \end{pmatrix}
  +\frac{\alpha-\beta-\gamma+\delta}{4}
  \begin{pmatrix} + & -\\ - & + \end{pmatrix}.
\end{align}
Now we can go back to the decision of fixing $b_1=1$ in order to make all coefficients positive.
For instance, if $\alpha+\beta-\gamma-\delta<0$ then we flip its sign as well as the one of the matrix it is in front of, which is allowed by the previous observation that $(a_1,a_2,b_1,b_2)$ and $(a_1,a_2,-b_1,-b_2)$ give opposite matrices.
This ensures the positivity of all coefficients.

The only missing ingredient to have a convex decomposition is that these coefficients should sum up to one.
In fact, as we can always complete with the point $\mathbf{0}$ which can be obtained by summing all vertices of \cref{eqn:ds22}, it is sufficient to have a sum less than or equal to one, that is,
\begin{equation}
  \sigma=\big|\alpha+\beta+\gamma+\delta\big|+\big|\alpha-\beta+\gamma-\delta\big|+\big|\alpha+\beta-\gamma-\delta\big|+\big|\alpha-\beta-\gamma+\delta\big|\leq4.
  \label{eqn:sigma}
\end{equation}
Note that this specific inequality was already given by Tsirelson in~\cite{Tsi87}.
Lemma~1 from~\cite{HQV+17} directly follows from the triangle inequality, which states that $\sigma\leq4(|\alpha|+|\beta|+|\gamma|+|\delta|)$ so that it is sufficient to enforce that the 1-norm is smaller than one to satisfy \cref{eqn:sigma}.
This inequality is, however, quite loose and can be improved thanks to the Cauchy--Schwarz inequality:
\begin{equation}
  \sigma^2\leq\left(1^2+1^2+1^2+1^2\right)\Big[\big(\alpha+\beta+\gamma+\delta\big)^2+\big(\alpha-\beta+\gamma-\delta\big)^2+\big(\alpha+\beta-\gamma-\delta\big)^2+\big(\alpha-\beta-\gamma+\delta\big)^2\Big]\leq16\big(\alpha^2+\beta^2+\gamma^2+\delta^2\big).
  \label{eqn:sigma2}
\end{equation}
For a matrix $\br$ with $\|\br\|_2\leq1$ we have $\alpha^2+\beta^2+\gamma^2+\delta^2\leq1$.
Then \cref{eqn:sigma2} implies that \cref{eqn:sigma} is satisfied, which concludes the proof for this small example.

\section{Multipartite scenarios}
\label{app:multipartite}

In this section, we generalise the main construction of local decomposition via Frank-Wolfe to multipartite states.
We prove the lemma used for the last analytical step and illustrated in \cref{app:ana}, and then extend it to include marginals.
We also provide details about the bounds in \cref{tab:tripartite} in the main text and give further analytical bounds for the GHZ state by exhibiting facets of the local polytope.

\subsection{Notation}

We consider a Bell scenario with $N$ parties receiving $m$ inputs and giving two outputs.
The parties are labelled by $n=1\ldots N$, the inputs by $x_n=1\ldots m$, and the outputs by $a_n=\pm1$.
Similarly to the bipartite case where the correlation matrix $\langle a_1a_2\rangle$ contains, together with the marginals $\langle a_1 \rangle$ and $\langle a_2 \rangle$, all the information about the probability distribution $p(a_1a_2|x_1x_2)$, we can encode the general multipartite correlation information in many tensors of orders ranging from $1$ to $N$.

Formally, we define
\begin{equation}
  r_{x_{n_1}\ldots x_{n_M}}^{(n_1\ldots n_M)}=\langle a_{n_1}\ldots a_{n_M}\rangle=\sum_{\hb{a_{n_1}\ldots a_{n_M}}{\pm1}}\bigg(\prod_{j\in[M]}a_{n_j}\bigg)p(a_{n_1}\ldots a_{n_M} | x_{n_1}\ldots x_{n_M}),
  \label{eqn:marginals}
\end{equation}
where the exponent $(n_1\ldots n_M)$ indicating the subset of $M$ parties considered will be dropped when this subset is clear from the indices $x_{n_1}\ldots x_{n_M}$.
Note that, in practice, one can simply work with a single correlation tensor of order $N$ and whose indices $x_n=0\ldots m$ incorporate both the tensors where $n$ appears (when $x_n=1\ldots m$) and those where it does not (when $x_n=0$).
The fixed coordinate $r_{0\ldots0}=1$ is omitted from all vectorised tensors and their norms.
With this, the dimension of the multipartite local polytope is $(m+1)^N-1$; note that the vertices of this polytope are the $2^{Nm}$ possible deterministic strategies.

A quantum correlation tensor is obtained when measuring the multipartite state $\rho$ with the dichotomic observables $A^{(n)}_{x_n}$, which gives, according to the Born rule,
\begin{equation}
  r_{x_1\ldots x_N}=\Tr\left[\left(A^{(1)}_{x_1}\otimes\ldots\otimes A^{(N)}_{x_N}\right)\rho\,\right],
  \label{eqn:quantum_correlation}
\end{equation}
with the convention that $A^{(n)}_0=\id$.

\subsection{Proof of \texorpdfstring{\cref{lem:ball}}{Lemma 1}}
\label{app:lemma}

Now we give the proof of our extension of~\cite[Lemma~1]{HQV+17}, namely, \cref{lem:ball} in \cref{app:ana}.
Here we consider only full correlators, that is, the tensor of order $N$ with $x_1,\ldots,x_N\in[m]$.
A corresponding criterion for full correlators of an $N$-qubit quantum system is given in~\cite[Eq.~(15)]{ZB02}; the proof below does not assume a quantum realisation.
Recall that, in this article, we always refer to the norms of the vectorised tensors that we consider.

To prove this, we consider a point $\br$ such that $\|\br\|_2\leq1$ and we exhibit an explicit decomposition over the deterministic strategies.
We start by writing it down in the standard basis:
\begin{equation}
  \br=\sum_{x_1\in[m]}\ldots\sum_{x_N\in[m]}r_{x_1\ldots x_N}\be_{x_1\ldots x_N}.
\end{equation}
Then we can decompose each of the $\be_{x_1\ldots x_N}$ in terms of deterministic strategies parametrised by $\va^{(1)}\ldots\va^{(N)}\in\{-1,+1\}^m$ and regroup the terms, namely,
\begin{equation*}
  \resizebox{\linewidth}{!}{$
    \br=\sum\limits_{x_1\ldots x_N}r_{x_1\ldots x_N}\left(\frac{1}{2^{Nm-1}}\sum\limits_{\va^{(1)}}\ldots\sum\limits_{\va^{(N)}}\frac{1+\prod\limits_{i\in[N]}a^{(i)}_{x_i}}{2}\,\bd_{\va^{(1)}\ldots\va^{(N)}}\right)=\frac{1}{2^{Nm-1}}\sum\limits_{\hb{\va^{(1)}\ldots\va^{(N)}}{}}\left(\sum\limits_{x_1\ldots x_N}r_{x_1\ldots x_N}\frac{1+\prod\limits_{i\in[N]}a^{(i)}_{x_i}}{2}\right)\bd_{\va^{(1)}\ldots\va^{(N)}},
  $}
\end{equation*}
where $\bd_{\va^{(1)}\ldots\va^{(N)}}$ also contains only full correlators.
Here we can split the sum over $\va^{(1)}\ldots\va^{(N)}$ into two parts depending on the sign of $\prod_{i\in[N]}a^{(i)}_1$.
In the sum where this last product is negative, we use the reindexing $\va^{(1)}\rightarrow-\va^{(1)}$, noting that $\bd_{\va^{(1)}\ldots\va^{(N)}}\rightarrow-\bd_{\va^{(1)}\ldots\va^{(N)}}$ and that $\prod_{i\in[N]}a^{(i)}_{x_i}\rightarrow-\prod_{i\in[N]}a^{(i)}_{x_i}$.
This allows us to regroup the two parts into one sum, that is,
\begin{align}
  \br
  =&\,\frac{1}{2^{Nm-1}}\sum_{\hb{\va^{(1)}\ldots\va^{(N)}}{\prod_{i\in[N]}a^{(i)}_1=1}}\overbrace{\sum_{x_1\ldots x_N}r_{x_1\ldots x_N}\prod_{i\in[N]}a^{(i)}_{x_i}}^{w_{\va^{(1)}\ldots\va^{(N)}}}\bd_{\va^{(1)}\ldots\va^{(N)}}=\frac{1}{2^{Nm-1}}\sum_{\hb{\va^{(1)}\ldots\va^{(N)}}{\prod_{i\in[N]}a^{(i)}_1=1}}\pm\big|w_{\va^{(1)}\ldots\va^{(N)}}\big|\,\bd_{\va^{(1)}\ldots\va^{(N)}}\\
  =&\sum_{\hb{\va^{(1)}\ldots\va^{(N)}}{\prod_{i\in[N]}a^{(i)}_1=1}}\frac{\big|w_{\va^{(1)}\ldots\va^{(N)}}\big|}{2^{Nm-1}}\,\bd_{\pm\va^{(1)}\va^{(2)}\ldots\va^{(N)}},\label{eqn:multidecomposition}
\end{align}
which turns out to be a valid local decomposition once we prove that the sum of its (nonnegative) coefficients is at most one.
Recall that we can indeed complete the decomposition with the zero correlation tensor obtained by averaging over all deterministic strategies.

In order to bound this sum, we apply the Cauchy--Schwarz inequality:
\begin{equation}
  \Bigg(\sum_{\hb{\va^{(1)}\ldots\va^{(N)}}{\prod_{i\in[N]}a^{(i)}_1=1}}\frac{\big|w_{\va^{(1)}\ldots\va^{(N)}}\big|}{2^{Nm-1}}\Bigg)^2
  \leq\sum_{\hb{\va^{(1)}\ldots\va^{(N)}}{\prod_{i\in[N]}a^{(i)}_1=1}}\left(\frac{1}{2^{Nm-1}}\right)^2\sum_{\hb{\va^{(1)}\ldots\va^{(N)}}{\prod_{i\in[N]}a^{(i)}_1=1}}\big|w_{\va^{(1)}\ldots\va^{(N)}}\big|^2=\|\br\|_2^2\leq1,
\end{equation}
where the last equality comes from the computation of all products appearing in the square of $w_{\va^{(1)}\ldots\va^{(N)}}$; all cross terms vanish.
More formally, the product of any two terms appearing in $w_{\va^{(1)}\ldots\va^{(N)}}$ can be computed as follows:
\begin{align}
  \sum_{\hb{\va^{(1)}\ldots\va^{(N)}}{\prod_{i\in[N]}a^{(i)}_1=1}}\bigg(r_{x_1\ldots x_N}\prod_{i\in[N]}a^{(i)}_{x_i}\bigg)\bigg(r_{x'_1\ldots x'_N}\prod_{i\in[N]}a^{(i)}_{x'_i}\bigg)
  =&\,r_{x_1\ldots x_N}r_{x'_1\ldots x'_N}\sum_{\hb{\va^{(1)}\ldots\va^{(N)}}{\prod_{i\in[N]}a^{(i)}_1=1}}\prod_{i\in[N]}a^{(i)}_{x_i}a^{(i)}_{x'_i}\nonumber\\
  =&\,r_{x_1\ldots x_N}^2\bigg(\prod_{i\in[N]}\delta_{x_i,x'_i}\bigg)2^{Nm-1},\label{eqn:wsquare}
\end{align}
where $\delta$ is the Kronecker delta function.
Indeed, flipping $\va^{(1)}$ preserves the product in the first line of \cref{eqn:wsquare} and exchanges the two signs of $\prod_{i\in[N]}a^{(i)}_1$, so the restricted sum is half the unrestricted one.
In the unrestricted sum, all signs are independent, and the average is zero unless $x_i=x'_i$ for every party.
This concludes the proof.

\subsection{Extension to marginals}
\label{app:lemma_marginals}

When marginals are included, flipping one party's outputs no longer negates the whole correlation tensor, so the sign changes used in \cref{eqn:multidecomposition} cannot be applied directly.
We can nevertheless use \cref{lem:ball} separately for each subset of parties to obtain the following result.
Here $\br^{(n_1\ldots n_M)}$ denotes the tensor with entries $r^{(n_1\ldots n_M)}_{x_{n_1}\ldots x_{n_M}}$ defined in \cref{eqn:marginals}.

\begin{lemma}
  A correlation tensor $\br$ including marginals is local whenever
  \begin{equation}
    \sum_{M\in[N]}\sum_{\hb{1\leq n_1<\ldots<n_M\leq N}{}}\big\|\br^{(n_1\ldots n_M)}\big\|_2\leq1.
    \label{eqn:ball_marginals}
  \end{equation}
  In particular, the closed ball of radius $1/\sqrt{2^N-1}$ for the 2-norm is contained in the local polytope including marginals.
  This radius is optimal.
  \label{lem:ball_marginals}
\end{lemma}

To prove this, fix a subset of parties $n_1,\ldots,n_M$ and suppose first that $\|\br^{(n_1\ldots n_M)}\|_2\leq1$.
By \cref{lem:ball}, this tensor has a local model when considered as a full-correlation tensor for these $M$ parties.
We can make all its proper marginals vanish by multiplying the outputs of party $n_j$ by a shared random sign $s_j$, the signs being uniformly distributed subject to $\prod_{j\in[M]}s_j=1$.
Indeed, this preserves the full correlators, whereas the average of every nonempty proper product of the signs is zero.
For $M=1$, the single sign is fixed to 1 and there are no proper marginals to remove.
We then let every party outside this subset output an independent unbiased sign.
This gives a local model whose only nonzero tensor is $\br^{(n_1\ldots n_M)}$.

For a general $\br$ satisfying \cref{eqn:ball_marginals}, we apply this construction to each nonzero tensor $\br^{(n_1\ldots n_M)}/\|\br^{(n_1\ldots n_M)}\|_2$ and mix the resulting local models with weights $\|\br^{(n_1\ldots n_M)}\|_2$.
As before, we complete the decomposition with the zero correlation tensor if the sum of the weights is less than one.
The claimed radius follows from the Cauchy--Schwarz inequality, since
\begin{equation}
  \sum_{M\in[N]}\sum_{\hb{1\leq n_1<\ldots<n_M\leq N}{}}\big\|\br^{(n_1\ldots n_M)}\big\|_2
  \leq\sqrt{2^N-1}\,\|\br\|_2.
\end{equation}
To see that the radius is optimal, take all the $2^N-1$ nonempty correlators at the input tuple $x_1=\ldots=x_N=1$ equal to $-t/(2^N-1)$ and all other correlators equal to zero.
The resulting tensor has norm $t/\sqrt{2^N-1}$, but $p(+1,\ldots,+1|1,\ldots,1)=(1-t)/2^N$ is negative for $t>1$, so this tensor is not local.
This concludes the proof.


\subsection{Results for tripartite GHZ and W states}

The tripartite GHZ and W states are defined as follows:
\begin{equation}
  \ket{\mathrm{GHZ}_3}=\frac{\ket{000}+\ket{111}}{\sqrt2}\quad\text{and}\quad\ket{\mathrm{W}_3}=\frac{\ket{001}+\ket{010}+\ket{100}}{\sqrt{3}}.
  \label{eqn:GHZ}
\end{equation}
Contrary to the singlet state $\ket{\psi^-}$ considered in the main text, these states have a privileged basis.

\paragraph{Lower bounds.}

The procedure to obtain lower bounds on the nonlocality threshold for multipartite states follows exactly the same lines as in the bipartite case, see \cref{fig:pipeline}.
We select a number of inputs $m$, choose a polyhedron on the Bloch sphere achieving a satisfying shrinking factor $\eta$, and run \cref{algo:bpcg} on the resulting correlation tensor (with marginals shrunk to anticipate on the last step).
By virtue of \cref{lem:ball_marginals}, an analyticity factor $\nu_2$ can be computed as in \cref{eqn:y} by choosing $R=1/\sqrt7$ with the Euclidean norm or by choosing $R=1$ with the norm from the left-hand size of \cref{eqn:ball_marginals}.
The subsequent shrinking step is used to obtain the following bound:
\begin{equation}
  v_c\geq\eta^N\nu_2v_0.
  \label{eqn:multilow}
\end{equation}
Note the exponent $N$ of $\eta$, accounting for the simulation of all projective measurements performed on each of the $N$ parties.

For the GHZ state, we use the polyhedron with $m=61$ and $\eta\approx0.9792$ given in \cref{tab:polyhedra} and choose the initial visibility $v_0=0.494$.
Note that we include marginals in the resulting correlation tensor $\bp$.
\cref{algo:bpcg} gives 54486 deterministic strategies decomposing $\bx_T$ such that we have $\nu_2\approx0.9997$.
We eventually obtain the bound $v_c^{\mathrm{GHZ}_3}\geq0.463679$ given in \cref{tab:tripartite}.

For the W state, we use the polyhedron with $m=61$ and $\eta\approx0.9792$ given in \cref{tab:polyhedra} and choose the initial visibility $v_0=0.524$.
Note that we include marginals in the resulting correlation tensor $\bp$.
\cref{algo:bpcg} gives 33529 deterministic strategies decomposing $\bx_T$ such that we have $\nu_2\approx0.9981$.
We eventually obtain the bound $v_c^{\mathrm{W}_3}\geq0.491051$ given in \cref{tab:tripartite}.

\paragraph{Upper bounds.}

For the tripartite GHZ state, following~\cite{BNV16}, we investigate the correlations obtained when performing measurements forming a regular polygon on the XY plane of the Bloch sphere, namely,
\begin{equation}
  A^{(n)}_{x_n}=\left(\cos\frac{(x_n-1)\pi}{m}\ ,\ \sin\frac{(x_n-1)\pi}{m}\ ,\ 0\right)\cdot\vec{\sigma}.
  \label{eqn:polygon}
\end{equation}
In this case, from \cref{eqn:quantum_correlation} we see that all marginals disappear and that the correlation tensor has elements
\begin{equation}
  r_{x_1x_2x_3}=\cos\frac{\left(x_1+x_2+x_3-3\right)\pi}{m}
  \label{eqn:GHZtensor}
\end{equation}
for $x_1,x_2,x_3\in[m]$; see also~\cite[Eq.~(50)]{BNV16}.

We can then run \cref{algo:bpcg} and extract a separating hyperplane from the output as in the bipartite case.
Moreover, this Bell inequality is actually a facet in this case, which allows us to derive the analytical expression of the nonlocality threshold for these measurements.
Note that, although it is possible to reformulate the cubic problem corresponding to finding the local bound of this Bell inequality as a QUBO problem~\cite{CELR22}, we enumerate all possible deterministic strategies here, as it is still possible in this case.
We eventually obtain the following exact threshold (recall that the Bell inequality is a facet):
\begin{equation}
  v_c^{\mathrm{GHZ}_3}\leq v_{16}^{\mathrm{GHZ}_3}=\mbox{\scalebox{0.97}{$\Big(\frac{1}{2}+\frac{\sqrt{2}}{3}-\frac{120\cos\frac\pi8+56\sin\frac\pi8}{137}+\frac{11960704\cos\frac\pi{16}-11459008\sin\frac\pi{16}}{9508049}+\frac{15419776\cos\frac{3\pi}{16}-5453760\sin\frac{3\pi}{16}}{9508049}\Big)^{-1}$}}\approx0.49160.
  \label{eqn:GHZ3_16}
\end{equation}

Since the set of all projective measurements on the XY plane is simulable by means of the ones from \cref{eqn:polygon} up to a shrinking factor $\eta=\cos[\pi/(2m)]$, we can use the local models obtained on the facet to derive lower bounds valid for the value $v_{XY}^{\mathrm{GHZ}_N}$ of the nonlocality threshold of the GHZ state under those planar measurements:
\begin{equation}
  v_{XY}^{\mathrm{GHZ}_3}\geq\cos\left(\frac{\pi}{32}\right)^3v_{16}^{\mathrm{GHZ}_3}\approx0.48453.
  \label{eqn:GHZ3_16_XY}
\end{equation}

For the tripartite W state, we choose measurements whose Bloch vectors form a pentakis dodecahedron having two vertices on the $Z$ axis.
We choose $v_0=0.5483$, extract a Bell inequality from the output of \cref{algo:bpcg} (not a facet here), and compute its local bound to obtain the bound presented in \cref{tab:tripartite}:
\begin{equation}
  v_c^{\mathrm{W}_3}\leq\mbox{$\frac{1000910150415}{314450816715+210987849189\sqrt5+\sqrt{565536694605181420689075+230287482340357705773330\sqrt5}}$}\approx 0.548236.
\end{equation}

\end{document}